\documentclass[10pt,conference]{IEEEtran}
\IEEEoverridecommandlockouts

\usepackage{cite}
\usepackage{amsmath,amssymb,amsfonts}
\usepackage{graphicx}
\usepackage{textcomp}
\usepackage{xcolor}

\usepackage{tikz}
\usepackage{amsmath}
\usepackage{mathrsfs} 
\usepackage{enumitem} 
\usepackage{cuted} 
\usepackage{booktabs} 
\usepackage{array} 
\usepackage[breakable]{tcolorbox} 
\tcbuselibrary{skins}
\usepackage{listings} 
\usepackage{listings-solidity}
\usepackage{algorithm}
\usepackage{algpseudocode}
\usepackage{xurl}

\definecolor{verylightgray}{rgb}{.97,.97,.97}
\definecolor{keywordblue}{rgb}{0.0,0.0,0.7}
\definecolor{commentgreen}{rgb}{0.25,0.5,0.35}
\definecolor{stringred}{rgb}{0.6,0.1,0.1}

\usepackage{tabularx} 
\usepackage{threeparttable} 

\def\BibTeX{{\rm B\kern-.05em{\sc i\kern-.025em b}\kern-.08em
    T\kern-.1667em\lower.7ex\hbox{E}\kern-.125emX}}
\DeclareRobustCommand{\KASS}{\ensuremath{\mathcal{KASS}}}
\begin{document}

\title{Beyond Detection: Agentic Attack Synthesis and Simulation for Smart Contracts}

\author{
\IEEEauthorblockN{Xianhao Zhang\textsuperscript{1},
Jing Sun\textsuperscript{2},
Zijian Zhang\textsuperscript{1,*},
Ye Liu\textsuperscript{1,*},
Zhe Hou\textsuperscript{3},
Jiaqi Gao\textsuperscript{1}, and
Yuqiang Sun\textsuperscript{4}}
\IEEEauthorblockA{\textsuperscript{1}Beijing Institute of Technology, Haidian, Beijing, China}
\IEEEauthorblockA{\textsuperscript{2}University of Auckland, Auckland, New Zealand \quad
\textsuperscript{3}Griffith University, Brisbane, Queensland, Australia}
\IEEEauthorblockA{\textsuperscript{4}Nanyang Technological University, Singapore}
\IEEEauthorblockA{zhangxh@bit.edu.cn, jing.sun@auckland.ac.nz, zhangzijian@bit.edu.cn, ye.liu@bit.edu.cn,\\
z.hou@griffith.edu.au, judgegao06@gmail.com, suny0056@e.ntu.edu.sg}
\thanks{\textsuperscript{*}Corresponding author: Zijian Zhang, Ye Liu.}
}

\maketitle

\begin{abstract}
Smart contract vulnerabilities pose severe financial risks, yet existing security tools largely stop at vulnerability detection, offering limited support for explaining whether reported flaws are exploitable, how attacks unfold, and what concrete damage they cause. To bridge this gap, we propose \KASS{} (Knowledge-Augmented Attack Synthesis and Simulation), a multi-agent framework for executable smart contract exploit verification. \KASS{} decomposes automated exploit generation into planning, generation, and testing stages, and integrates three complementary mechanisms: retrieval-augmented planning over real-world audit knowledge, formal generation and validation constraints that bind attack plans to executable proof-of-concept tests, and a hierarchical dual-loop refinement process that repairs code-level errors while triggering strategy-level replanning when attack assumptions fail.
We evaluate \KASS{} on 104 SmartBugs-Curated contracts across four vulnerability categories. Experimental results show that \KASS{} successfully generates executable exploits for 94.23\% of tested contracts; this rate is higher than previously reported results for REX and AdvSCanner on comparable SmartBugs-Curated subsets, and higher than our reproduced Claude Code baseline under the same evaluation protocol.
On 11 real-world CVE-tagged contracts, \KASS{} successfully validates 9 cases. Beyond exploit generation, \KASS{} produces structured attack plans that document exploitation flows, quantify potential asset losses, and serve as semantic false positive filters for static analysis tools.
\end{abstract}


\section{Introduction}
\label{sec:introduction}

Smart contracts serve as the cornerstone of the Decentralized Finance (DeFi)~\cite{zetzsche2020decentralized} ecosystem. DeFi protocols managed more than US\$69 billion in total value locked (TVL) in 2026~\cite{defillama2026tvl}, yet exploit losses reached US\$512 million in 2025~\cite{hacken2025report}.
These losses persist despite mature traditional detectors~\cite{feist2019slither, mythril2026, mossberg2019manticore}, as well as newer LLM-based auditors that can identify complex, context-dependent vulnerabilities at scale~\cite{sun2024gptscan,wei2025smartauditflow, wei2025advanced}.

A central limitation of current smart contract security practice is that vulnerability detection is often treated as the end point of analysis. Audit reports, static analyzers, and emerging LLM-based auditors can identify suspicious code patterns or potential weaknesses, but these findings are not always validated through executable attacks. Consequently, many detected vulnerabilities remain textual claims or static warnings, without clear evidence of whether they can be triggered in practice, what protocol state or transaction sequence is required, and what concrete impact an attacker could achieve~\cite{zhang2023demystifying, chen2020survey, chaliasos2024smart}.

This detection-to-execution gap is particularly important because, among 4,364 contracts reported as vulnerable, roughly 75\% were considered unexploitable, corresponding to false positives or issues that did not constitute practical security risks~\cite{hu2024smart}.
Such cases may reflect methodological limitations, incomplete coverage, or the inherent complexity of attack surfaces.
They also point to a practical problem: identifying a potential weakness does not necessarily demonstrate its exploitability. Without executable validation, it remains difficult to distinguish false positives, theoretical issues, and low-impact findings from vulnerabilities that can lead to serious protocol compromise.

Automated Exploit Generation (AEG) offers a promising way to close this gap by moving beyond vulnerability detection toward executable exploit realization. Given a detected smart contract vulnerability, an AEG system should automatically generate a feasible attack scenario, instantiate the required protocol state and transaction sequence, implement the attack as a reproducible test case, and execute it in a controlled environment to measure practical consequences. However, existing AEG techniques remain limited. Prior methods~\cite{krupp2018teether,wang2020oracle,wu2024advscanner,xiao2025prompt,claude-code} often rely on predefined vulnerability assumptions, templates, or oracles; struggle to reconstruct complex initialization logic, protocol state, and multi-contract interactions; and may generate test harnesses that compile but do not actually exercise the vulnerable path or quantify concrete attack impact. As a result, they provide only partial support for turning detected vulnerabilities into reproducible, semantically validated exploits.

Building on this motivation, we propose \KASS{} (Knowledge-Augmented Attack Synthesis and Simulation) as an execution-based framework for smart contract security assessment. Rather than stopping at vulnerability reports, \KASS{} generates feasible attack scenarios, reconstructs the required contract state and function interaction sequence, and realizes them as reproducible Foundry test cases. To make this process reliable, \KASS{} combines formalized constraints on exploit objectives and expected post-attack effects with loop-engineering mechanisms~\cite{loop-engineering} that iteratively refine generated tests and revisit attack assumptions when execution fails. Executing these tests allows \KASS{} to determine whether a reported flaw can be concretely exploited and to measure its practical consequences, thereby providing stronger exploitability evidence and helping prioritize security findings.

\begin{figure*}[h]
  \centering
  \includegraphics[width=0.80\linewidth]{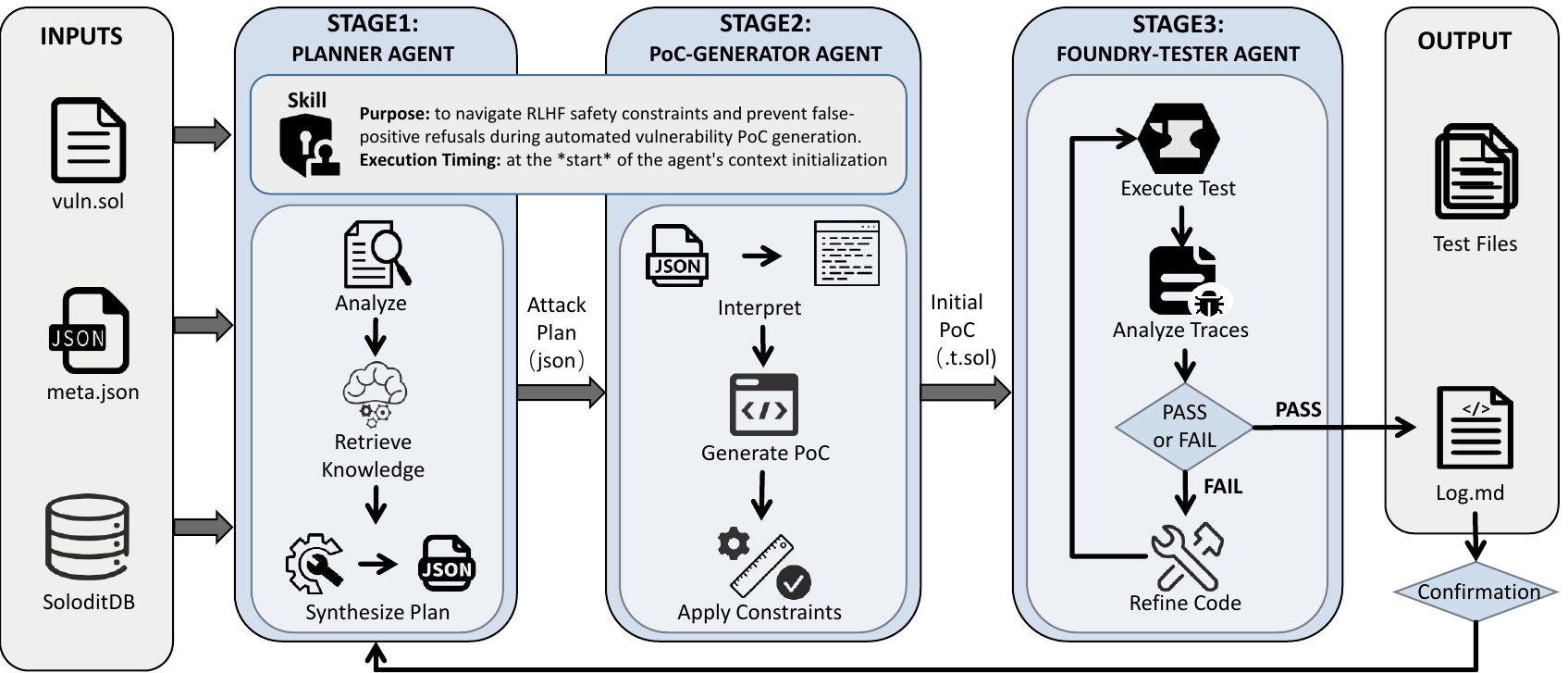}
  \caption{Overview of the Knowledge-Augmented Attack Synthesis and Simulation framework.
The system operates through a hierarchical dual-loop mechanism: an inner loop for code-level refinement (Stage 3) and an outer loop for strategy-level replanning (Stage 1), grounded by an external vulnerability knowledge base.}
  \label{fig:overview}
\end{figure*}

Our main contributions are as follows:
\begin{itemize}[leftmargin=*, nosep]
    \item We propose \KASS{}, a multi-agent framework that bridges vulnerability detection and executable exploit verification by decomposing smart contract AEG into planning, generation, and testing stages.
    \item We design three complementary mechanisms---retrieval-augmented planning over real-world audit knowledge, formal generation and validation constraints for executable PoCs, and hierarchical dual-loop refinement for both code-level repair and strategy-level replanning---to transform vulnerability reports into concrete exploit objectives, vulnerable-path tests, and validated attack outcomes.
    \item We evaluate \KASS{} on benchmark and real-world contracts, showing a 94.23\% success rate on SmartBugs-Curated, higher than previously reported REX and AdvSCanner results on comparable subsets and higher than our same-protocol Claude Code baseline, while validating 9 of 11 CVE-tagged cases and supporting damage quantification and semantic false positive filtering.
\end{itemize}

\noindent Paper Organization. Section~\ref{sec:background} introduces background on smart contracts and automated exploit generation. Section~\ref{sec:methodology} presents the \KASS{} framework and its three-agent architecture. Section~\ref{sec:evaluation} describes our experimental setup and answers the five research questions. Section~\ref{sec:discuss} discusses threats to validity and reviews related work, and Section~\ref{sec:conclusion} concludes.

\section{Background}
\label{sec:background}

\subsection{Smart Contract}

A smart contract is a blockchain-deployed program that automatically enforces predefined rules without intermediaries.

\textbf{Vulnerabilities.} Since the inception of Ethereum~\cite{buterin2013ethereum}, the proliferation of deployed smart contracts has introduced numerous security issues. The most widely adopted vulnerability taxonomies are DASP (Decentralized Application Security Project)~\cite{dasp2018} and SWC (Smart Contract Weakness Classification)~\cite{swcregistry}, which categorize common patterns. Zhang et al.~\cite{zhang2023demystifying} further classify vulnerabilities into machine-auditable bugs (e.g., reentrancy, integer overflow) that can be detected by automated tools, and machine-unauditable bugs (e.g., price oracle manipulation) that require domain-specific knowledge.

\textbf{Testing Suite.} Foundry~\cite{foundry} is a smart contract development toolkit that has become the de facto standard for Ethereum security testing. Its testing framework supports Solidity-native tests and provides cheatcodes for fine-grained blockchain state manipulation, making it well-suited for simulating complex attack scenarios. In \KASS{}, we leverage Foundry as the execution backend for validating generated exploits.

\subsection{Automated Exploit Generation}

Automated Exploit Generation refers to a class of techniques designed to automatically generate functional exploits for given vulnerabilities. This concept was famously introduced and popularized by Avgerinos et al.~\cite{avgerinos2014automatic}, typically combining automated techniques such as symbolic execution, dynamic analysis, or genetic algorithms. Unlike fuzzing, which primarily focuses on discovering program crashes or the existence of vulnerabilities, the core objective of AEG is to generate an exploit or Proof-of-Concept (PoC) that verifies the vulnerability is in an \textit{exploitable state}.

The majority of existing AEG research targets memory-based vulnerabilities in C/C++ programs via control-flow hijacking or data-oriented techniques~\cite{bui2025systematic, avgerinos2014automatic, schwartz2011q}.
For smart contracts, AEG instead synthesizes transaction sequences that trigger vulnerable states and demonstrate concrete exploit effects under complex on-chain state and inter-contract dependencies.

\section{Methodology}
\label{sec:methodology}

This section presents the \KASS{} framework, including its overall architecture, and three core agents.

\subsection{Framework Overview}

As illustrated in Figure~\ref{fig:overview}, \KASS{} is organized as a loop-engineered multi-agent workflow for smart contract exploit generation. Given a vulnerable contract (\texttt{vuln.sol}), vulnerability metadata (\texttt{meta.json}), and access to the Solodit knowledge base, the framework coordinates three agents around a shared objective: converting a vulnerability report into an executable Foundry proof of exploit.

The three agents collaborate through explicit intermediate artifacts rather than a single end-to-end prompt:

\begin{itemize}[leftmargin=*, nosep]
  \item \textbf{Planner Agent.} Retrieves relevant audit findings and converts them, together with the target contract logic, into a structured Attack Plan that specifies the exploit objective, preconditions, interaction sequence, and success oracle.
  \item \textbf{PoC-Generator Agent.} Translates the Attack Plan into executable Solidity artifacts, including the attack contract and Foundry test harness, while enforcing version, interface, and oracle constraints.
  \item \textbf{Foundry-Tester Agent.} Executes the generated PoC, inspects compiler/runtime feedback and transaction traces, and determines whether the exploit path satisfies the specified postconditions.
\end{itemize}

The design has three key elements, each tied to one agent. First, the Planner Agent integrates external security knowledge so that attack planning is grounded in real audit evidence.
Second, the PoC-Generator Agent binds the attack plan to formal validation constraints, which improves generation quality by encouraging executable interaction sequences and explicit post-attack checks.
Third, the Foundry-Tester Agent implements the hierarchical dual loop: local implementation repair in the inner loop and strategy-level replanning in the outer loop.

Both the Planner and PoC-Generator agents are initialized with safety-navigation skills that establish a professional security-auditor context and restrict reasoning to public, authorized artifacts. The final output consists of the generated exploit files and a reproducible execution log (\texttt{Log.md}) documenting the planning, generation, testing, and refinement process.

\subsection{Knowledge-Augmented Reasoner: The Planner Agent}

The objective of the first stage is to bridge the abstraction-implementation gap, i.e., translating a high-level vulnerability classification into a precise, executable sequence of on-chain interactions. To achieve this, we deploy the Planner Agent, a specialized reasoning unit designed to function as a retrieval-augmented Bayesian inference engine.

\subsubsection{Formal Agent Workflow}

We model the planning process by defining the Planner Agent as a mapping function $F_{plan}$ that transforms raw vulnerability data into a structured attack plan. The execution logic follows a three-step protocol:

\begin{enumerate}[leftmargin=*, nosep]
    \item Contextual Analysis ($C_{local}$):
    Given the source code $C_{vul}$ and vulnerability metadata $v$ (containing location $v_{loc}$ and type $v_{type}$), the agent extracts a Local Dependency Context $C_{local}$:
    \begin{equation}
        C_{local} = \text{Extract}(C_{vul}, v_{loc})
    \end{equation}
    where Extract uses $v_{loc}$ as an anchor to traverse the call graph of $C_{vul}$, collecting all state variables, function bodies, modifiers, and access control checks that have a transitive data or control dependency on the vulnerability site. This discards irrelevant contract logic and retains only the elements necessary to reason about triggering the vulnerability.

    \item Knowledge Retrieval ($S_{ret}$):
    To ground the reasoning in real-world exploitation evidence, the agent issues a keyword query to the external knowledge base $\mathcal{K}$ (Solodit) via its API, using $v_{type}$ as the search key:
    \begin{equation}
        S_{ret} = \text{Retrieve}(v_{type}, \mathcal{K})
    \end{equation}
    where Retrieve returns the top-$k$ matching audit reports, ranked by relevance to $v_{type}$. Each report in $S_{ret}$ provides real-world exploitation examples, known attack patterns, and analogous vulnerability instances from production contracts.

    \item Structured Plan Synthesis ($\pi$):
    Finally, the agent combines $C_{local}$ and $S_{ret}$ to produce the attack plan $\pi$:
    \begin{equation}
        \pi = \text{Synthesize}(C_{local}, S_{ret}) \in \mathbb{S}
    \end{equation}
    where Synthesize is an LLM-conditioned generation step: the model $\mathcal{M}$ receives a structured prompt template $\Psi$ that concatenates $C_{local}$ and $S_{ret}$, producing $\pi = \mathcal{M}(\Psi(C_{local}, S_{ret}))$. The output is constrained to the schema $\mathbb{S}$ via structured JSON decoding, ensuring $\pi$ is a machine-parseable blueprint rather than free-form text.

    The schema $\mathbb{S}$ requires that $\pi$ be a JSON object containing five semantic fields (detailed in Table~\ref{tab:attack-plan-fields}):
    \begin{equation}
        \mathbb{S} = \{ \text{Explanation}, P_{prep}, I_{nt}, S_{post}, O_{bj} \}
    \end{equation}
\end{enumerate}

Combining these three steps, the complete planning function can be compactly expressed as:
\begin{equation}
    \pi = F_{plan}(C_{vul}, v, \mathcal{K})
\end{equation}

\begin{table}[t]
\centering
\caption{Semantic fields of the structured attack plan.}
\label{tab:attack-plan-fields}
\small
\renewcommand{\arraystretch}{1.4}
\setlength{\tabcolsep}{4pt}
\begin{tabular}{@{} p{2cm} >{\centering\arraybackslash}p{1cm} p{4cm} @{}}
\toprule
\textbf{Field} & \textbf{Symbol} & \textbf{Description} \\
\midrule
Explanation & -- & Root cause analysis linking $v_{type}$ to the logic in $C_{vul}$ \\[2pt]
Preparation & $P_{prep}$ & Pre-conditions such as flash loan acquisition or contract deployment \\[2pt]
Interaction & $I_{nt}$ & Ordered function call sequence (e.g., \texttt{deposit} $\to$ \texttt{reenter} $\to$ \texttt{withdraw}) \\[2pt]
Post-State & $S_{post}$ & Quantifiable success condition \\[2pt]
Objective & $O_{bj}$ & Categorical strategic goal \\
\bottomrule
\end{tabular}
\end{table}

The five-field schema follows established attack-modeling principles. Inspired by the Cyber Kill Chain~\cite{hutchins2011intelligence}, it separates target understanding (\textit{Explanation}), exploit setup and execution (\textit{Preparation} and \textit{Interaction}), and success verification (\textit{Post-State} and \textit{Objective}). This structure also matches the preparation--interaction--outcome pattern observed in real-world DeFi exploits~\cite{zhou2023sok}.
Our structured plan is designed to encode this empirically validated attack lifecycle, so that the generated plans align with the operational structure observed in real-world exploits rather than being constructed ad hoc.
To prevent the Planner from producing vague or non-verifiable objectives, \KASS{} further restricts the \textit{Objective} field to four explicit categories: (a) \textit{Primary Financial Gain}, (b) \textit{Strategic Financial Positioning}, (c) \textit{Disruption/Sabotage/DoS}, and (d) \textit{Manipulation of System Behavior}.
This categorical constraint forces each plan to state the intended impact of the exploit in a form that can be translated into concrete post-state assertions during testing.

\subsubsection{Theoretical Justification: Retrieval as Bayesian Inference}

We model in-context learning as implicit Bayesian inference to explain why retrieval improves direct generation.

Let $\theta$ denote the latent exploitation logic, i.e., the sequence required to trigger the vulnerability. Without retrieved context, the model conditions on the target contract $C_{vul}$ and vulnerability type $v_{type}$, but lacks a concrete attack instantiation. The plan $\pi$ is therefore governed by the conditional prior:

\begin{equation}
    P_{\text{no-ret}}(\pi) = \int p(\pi \mid C_{vul}, \theta) \, p(\theta \mid v_{type}) \, d\theta
\end{equation}

Although $v_{type}$ (e.g., Reentrancy) narrows the search space, $p(\theta \mid v_{type})$ remains diffuse and multi-modal. Without concrete references, the model must infer variants such as cross-function or read-only reentrancy from the broad category alone, often producing generic or hallucinated patterns.

The Planner Agent retrieves a context set $S_{ret}$ from Solodit. These examples act as observed evidence that sharpens inference of $\theta$, so generation follows the posterior predictive distribution:

\begin{multline}
    P_{\text{planner}}(\pi \mid S_{ret}, C_{vul}) \propto \\
    \int p(\pi \mid C_{vul}, \theta) \underbrace{p(S_{ret} \mid \theta, v_{type})}_{\text{Likelihood}} p(\theta \mid v_{type}) \, d\theta
\end{multline}

Here, $p(S_{ret} \mid \theta, v_{type})$ is the likelihood that retrieved reports with rich attack trajectories are consistent with $\theta$. As shown by Xie et al.~\cite{xie2022icl}, longer in-context examples yield higher signal-to-noise ratios for inferring latent concepts. Since Solodit reports satisfy this condition, the likelihood term dominates the integral.

In effect, retrieval acts as a Bayesian update that collapses the broad prior $p(\theta \mid v_{type})$ into a sharper posterior around the optimal logic $\theta^*$.
This does not guarantee optimality; it biases the agent toward plans grounded in documented exploitation cases rather than unconstrained guesses.

\subsection{Syntax Translator: The PoC-Generator Agent}
\label{subsec:generater}

While the Planner Agent defines the \textit{semantic trajectory} of the exploit, its output remains an abstract JSON description. The objective of Stage 2 is to bridge the semantic-syntactic gap by transforming this high-level plan into executable code. We deploy the PoC-Generator Agent, a syntax-constrained translator that maps the logical steps in $\pi$ to a compilable Foundry test case.

\subsubsection{Formal Translation Process}
\label{subsec:formal-generator}

Let $\Pi$ be the space of valid attack plans and $\mathcal{T}$ be the space of valid Foundry test contracts. The generator defines a translation function $F_{gen}: \mathcal{C} \times \Pi \to \mathcal{T}$. Given the vulnerable contract $C_{vul}$ and the plan $\pi$, the agent generates an initial attack contract $C_{att}^{(0)}$:

\begin{equation}
    C_{att}^{(0)} = F_{gen}(C_{vul}, \pi)
\end{equation}

To increase the likelihood that $C_{att}^{(0)}$ is a functionally viable test case rather than hallucinatory text, we embed a set of syntactic constraints into the agent's system prompt. These constraints formally restrict the output space as follows.

\begin{itemize}[leftmargin=*]
    \item Version Compatibility ($\mathcal{C}_{ver}$). The prompt enforces \texttt{pragma} consistency, requiring the compiler version of the test to match the target:
    \begin{equation}
        Ver(C_{att}^{(0)}) \equiv Ver(C_{vul})
    \end{equation}
    This prevents compilation errors caused by breaking changes across Solidity versions (e.g., the deprecation of \texttt{SafeMath} in v0.8 due to built-in overflow checks~\cite{chen2020survey}).

    \item Minimalism ($\mathcal{C}_{min}$). To reduce noise and potential side effects, the agent is constrained to avoid importing unnecessary contracts. This restricts the dependency graph $D(C_{att}^{(0)})$ to the minimal set required to interface with $C_{vul}$:
    \begin{equation}
        D(C_{att}^{(0)}) \subseteq \{C_{vul}, \text{FoundryStd}\}
    \end{equation}

    \item Oracle Embedding ($\mathcal{C}_{oracle}$). The agent must generate a verifiable success condition by translating the natural language descriptions in $\pi.O_{bj}$ and $\pi.S_{post}$ into a boolean predicate $\phi(s)$ encoded as the final \texttt{require} statement:
    \begin{equation}
        \phi(s_{final}) =
        \begin{cases}
        \text{true}, & \text{if } \pi.O_{bj} \text{ is satisfied} \\
        \text{revert}, & \text{otherwise}
        \end{cases}
    \end{equation}
    This predicate $\phi$ serves as the definitive oracle for the subsequent testing stage.
\end{itemize}

\begin{figure}[t]
  \centering
  \includegraphics[width=0.95\linewidth]{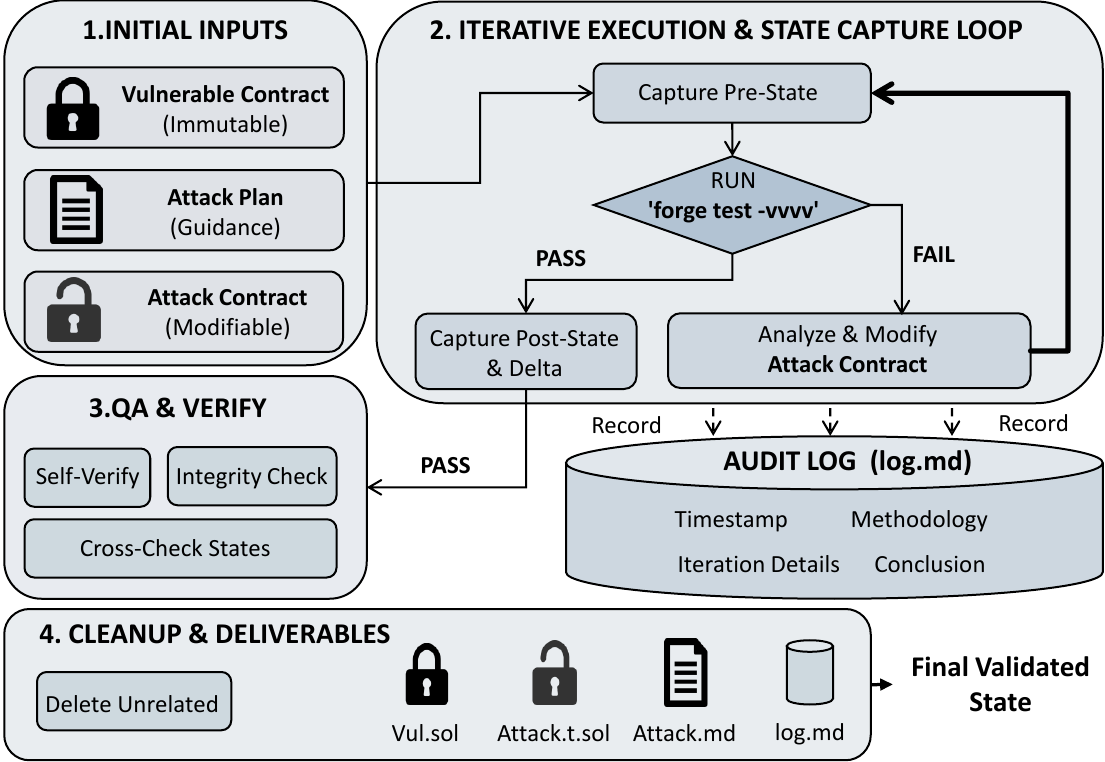}
  \caption{Workflow of the Foundry-Tester Agent. The inner loop iteratively refines code-level errors based on execution traces, while the outer loop triggers strategic replanning when the current approach is deemed inviable.}
  \label{fig:test-agent}
\end{figure}

\subsubsection{Structure-Aware Code Generation}

The agent implements the translation by mapping the semantic fields of $\pi$ to specific structural components of the Foundry test file (\texttt{.t.sol}):

\begin{itemize}[leftmargin=*]
    \item $\pi.P_{prep} \to$ \texttt{setUp()}. The preparation phase of the plan is translated into the \texttt{setUp()} function, which includes deploying the target contract $C_{vul}$ and utilizing Foundry cheatcodes such as \texttt{deal} and \texttt{prank} to establish the initial environment $s_0$.
    \item $\pi.I_{nt} \to$ \texttt{testExploit()}. The interaction sequence is translated into the primary test function. The agent converts logical steps (e.g., borrow flash loan, invoke vulnerable function, withdraw funds) into precise Solidity function calls with encoded parameters.
    \item Ambiguity Resolution. When explicit parameter values are missing in $\pi$, the agent leverages its parametric knowledge of common vulnerability patterns~\cite{chen2020survey} to infer reasonable defaults (e.g., assuming a standard ERC20 \texttt{approve} amount if unspecified).
\end{itemize}

The output of this stage consists of two artifacts: a Test Contract (\texttt{<vuln>\_Attack.t.sol}), which is a Solidity file implementing $C_{att}^{(0)}$ that imports and interfaces with the vulnerable contract $C_{vul}$; and an Execution Documentation (\texttt{attack.md}), a synthesized document explaining the mapping from $\pi$ to $C_{att}^{(0)}$, which aids interpretability and serves as a reference for the refinement loop in Stage~3.

By enforcing these syntactic constraints, the PoC-Generator Agent significantly increases the probability that the resulting code is locally compilable and semantically aligned with the strategic plan, providing a high-quality initialization for the iterative testing phase.

\subsection{Feedback-Driven Optimizer: The Foundry-Tester Agent}
\label{subsec:tester}

The final stage of the \KASS{} framework transitions from static generation to dynamic verification. We deploy the Foundry-Tester Agent, a specialized optimization unit operating within a Foundry environment.
As illustrated in Figure~\ref{fig:test-agent}, this agent operates through a dual-loop optimization architecture: an \textit{inner loop} that iteratively refines code-level errors (e.g., compilation failures, runtime reverts) based on execution traces, and an \textit{outer loop} that triggers strategic replanning when the current attack approach is deemed fundamentally inviable. The inner loop executes the generated PoC, captures state transitions, and refines the exploit code until the predefined success condition is satisfied or a local maximum is reached. When the inner loop exhausts its refinement budget without success, the outer loop escalates the failure to the preceding stages, requesting a revised attack plan $\pi'$ from the Attack-Planner Agent before reinitializing the code generation and testing pipeline.

\subsubsection{Formal Algorithmic Description}

\begin{algorithm}[t]
\caption{Iterative Exploit Refinement}
\label{alg:refine}
\small
\begin{algorithmic}[1]
\Require $C_{vul}$, $C_{att}^{(0)}$, Instruction $I$
\State $H_0 \leftarrow \emptyset$
\For{$t = 1$ \textbf{to} $T_{max}$}
    \State $res, \tau_t \leftarrow \text{Execute}(C_{att}^{(t-1)}, C_{vul})$
    \If{$res = \text{True}$}
        \State $\text{Cleanup}()$
        \State \Return $C_{att}^{(t-1)}$, $H_{t-1}$
    \Else
        \State $H_t \leftarrow H_{t-1} \cup \{(C_{att}^{(t-1)}, \tau_t)\}$
        \State $C_{att}^{(t)} \leftarrow \text{Refine}(C_{att}^{(t-1)}, \tau_t, I)$
    \EndIf
\EndFor
\State \Return Failure
\end{algorithmic}
\end{algorithm}

Algorithm~\ref{alg:refine} formalizes the iterative refinement process of the Foundry-Tester Agent. The algorithm takes three inputs: the vulnerable contract $C_{vul}$ (treated as read-only throughout the entire process), the initial attack contract $C_{att}^{(0)}$ generated by Stage~2, and an instruction document $I$ (i.e., \texttt{attack.md} and system prompt) that encodes the semantic mapping from the attack plan $\pi$ to executable code. The history buffer $H_t$ accumulates all prior attempts and their corresponding execution traces, providing the agent with an expanding context of previously observed failures. The agent operates through the following phases:

\begin{enumerate}[leftmargin=*, nosep]
    \item Initial Setup. The agent begins by reading the instruction document $I$ to internalize the expected exploit methodology. It examines both $C_{vul}$ and $C_{att}^{(0)}$, verifying interface compatibility and pragma consistency. A log file (\texttt{Log.md}) is initialized to record the timestamp, contract identifiers, and a summary of the attack strategy.

    \item Iterative Test-Refine Loop (Lines 3--10). At each iteration $t$, the agent invokes the Foundry execution engine via \texttt{forge test -vvvv} to obtain a binary result $res$ and a detailed execution trace $\tau_t$. The trace $\tau_t$ captures the full transaction call stack, including function invocations, state variable mutations, revert messages, and gas consumption. 
    The verbose trace output is parsed to classify the failure into actionable categories: compilation errors, runtime reverts, or assertion failures. Based on this classification, the agent modifies \emph{only} the attack contract $C_{att}$ while preserving $C_{vul}$ in its original state. Each modification and its rationale are documented in \texttt{Log.md} before re-execution.

    \item Success Determination (Lines 4--6). The result $res$ is determined to be True only when three conditions are jointly satisfied: (1)~the syntactic constraints from Stage~2 ($\mathcal{C}_{ver}$, $\mathcal{C}_{min}$, $\mathcal{C}_{oracle}$) are preserved in the refined code, (2)~the vulnerable contract $C_{vul}$ remains unmodified, and (3)~\texttt{forge test -vvvv} reports a pass.
For each test execution, the agent records the attacker's state before and after the attack, including token balances, contract storage, and ETH holdings, to quantify the exploit's impact. Upon a successful test pass, the agent performs one additional confirmation run to ensure result consistency.

    \item Artifact Cleanup (Line 5). The $\text{Cleanup}()$ function is invoked upon successful verification.
It produces four core output files: the original $C_{vul}$, the final refined $C_{att}^{(t^*)}$, the instruction document $I$, and a complete execution log \texttt{Log.md}. Required testing components are also retained, while redundant intermediate artifacts generated during testing are removed to maintain a clean workspace.
When $res = \text{True}$, the agent returns the verified attack contract along with the complete history $H_{t-1}$ for audit logging.
\end{enumerate}

\subsubsection{Intuition: Search Space Pruning via Execution Feedback}

The iterative refinement loop can be understood through the lens of \textit{Counterexample-Guided Inductive Synthesis} (CEGIS)~\cite{solar2006combinatorial}. Let $\Omega$ denote the space of all possible attack contracts. At each iteration $t$, the agent executes a candidate $C^{(t)}$ and observes an execution trace $\tau_t$. If the test fails, the trace identifies a subset $\Omega_{invalid} \subset \Omega$ of programs that would produce the same error, effectively pruning them from consideration.

Following the information-theoretic perspective on software testing~\cite{bohme2020boosting}, we quantify the progress of this search using information gain. Let $H(\mathcal{C}^* \mid \mathcal{H}_{t})$ denote the remaining uncertainty about the correct exploit $\mathcal{C}^*$ given the execution history $\mathcal{H}_t$. The information gain from trace $\tau_t$ is:
\begin{equation}
    IG(\mathcal{C}^*; \tau_t) = H(\mathcal{C}^* \mid \mathcal{H}_{t-1}) - H(\mathcal{C}^* \mid \mathcal{H}_{t-1}, \tau_t)
\end{equation}

Each informative trace yields positive information gain, monotonically reducing the entropy of the search space. This feedback is appended to the execution history and included in the prompt context, guiding the LLM to avoid previously observed failure patterns. Unlike stochastic fuzzing, this CEGIS-like mechanism transforms exploit synthesis into a directed search that converges toward the feasible solution.

\section{Evaluation}
\label{sec:evaluation}

In this section, we first describe our experimental setup, including research questions, dataset selection, baseline tools, evaluation metric and implementation details. We then present experimental results to answer five research questions.

\subsection{Experimental Setup}
\subsubsection{Research Questions}
We aim to answer the following research questions:
\begin{itemize}[leftmargin=*]
  \item \textbf{RQ1 (Intrinsic LLM Limitation)}: How well can LLMs perform AEG tasks relying solely on their pre-trained knowledge?
  \item \textbf{RQ2 (Effectiveness \& Efficiency)}: How effective and efficient is \KASS{} in automated exploit generation tasks compared to state-of-the-art tools?
  \item \textbf{RQ3 (Ablation \& Sensitivity Analysis)}: How does each component of \KASS{} contribute to overall performance, and how does the choice of LLM backbone affect the results?
  \item \textbf{RQ4 (Real-World CVE Validation)}: How does \KASS{} perform on real-world vulnerable smart contracts curated from CVE disclosures?
  \item \textbf{RQ5 (Exploitability Assessment)}: Can \KASS{}'s structured outputs support damage quantification and reduce false positives from static detection?
\end{itemize}

\subsubsection{Dataset}

SmartBugs-Curated~\cite{durieux2020empirical} is a widely-adopted benchmark in smart contract security research, comprising 10 sub-datasets with detailed vulnerability location annotations. We select four deterministic and clearly labeled categories totaling 104 contracts: Reentrancy~(31 samples, avg.\ 44 LOC), Arithmetic~(15 samples, avg.\ 95 LOC), Unchecked Low Level Calls~(52 samples, avg.\ 130 LOC), and Denial of Service~(6 samples, avg.\ 49 LOC).

To evaluate generalization beyond academic benchmarks, we searched public CVE databases for Solidity smart-contract vulnerabilities and manually retained 11 most recent cases with clear vulnerability descriptions and available matching source code. These CVE cases cover diverse contract types and code complexity, detailed information is reported in Table~\ref{tab:cve-results}.

\begin{table*}[htbp]
\centering
\small
\begin{threeparttable}
\caption{Comparison of Exploit Success Rates Across Vulnerability Types}
\label{tab:comparison}
\setlength{\tabcolsep}{4pt}
\begin{tabular}{@{}l*{5}{rr}@{}}
\toprule
& \multicolumn{2}{c}{\textbf{Reentrancy}} & \multicolumn{2}{c}{\textbf{Arithmetic}} & \multicolumn{2}{c}{\textbf{DoS}} & \multicolumn{2}{c}{\textbf{Unchecked}} & \multicolumn{2}{c}{\textbf{Total}} \\
\cmidrule(lr){2-3} \cmidrule(lr){4-5} \cmidrule(lr){6-7} \cmidrule(lr){8-9} \cmidrule(lr){10-11}
\textbf{Method} & \#Succ & Rate & \#Succ & Rate & \#Succ & Rate & \#Succ & Rate & \#Succ & Rate \\
\midrule
\multicolumn{11}{@{}l}{\textit{Pure Prompting}} \\
\quad GPT-5.1      & 4  & 12.90\% & 11 & 73.33\% & 3 & 50.00\% & 17 & 32.69\% & 35 & 33.65\% \\
\quad Gemini-3     & 0  & 0.00\%  & 1  & 6.67\%  & 1 & 16.67\% & 3  & 5.77\%  & 5  & 4.81\%  \\
\quad DeepSeek-V3.2& 2  & 6.45\%  & 3  & 20.00\% & 1 & 16.67\% & 4  & 7.69\%  & 10 & 9.62\%  \\
\midrule
\multicolumn{11}{@{}l}{\textit{AEG and Agentic Baselines}} \\
\quad REX          & 18 & 58.06\% & 13 & 86.67\% & 4 & 66.67\% & 17 & 32.69\% & 52 & 50.00\% \\
\quad AdvSCanner   & -- & 80\% & -- & --      & --& --      & -- & --      & -- & --      \\
\quad Claude Code & 13 & 41.94\% & 7 & 46.67\% & 1 & 16.67\% & 0 & 0.00\% & 21 & 20.19\% \\
\midrule
\multicolumn{11}{@{}l}{\textit{\KASS{}}} \\
\quad w/ GPT-5.1       & 29 & \textbf{93.55\%} & 15 & \textbf{100.00\%} & 5 & \textbf{83.33\%} & 49 & \textbf{94.23\%} & 98 & \textbf{94.23\%} \\
\quad w/ Gemini-3-Flash      & 25 & 80.65\% & 13 & 86.67\% & 3 & 50.00\% & 47 & 90.38\% & 88 & 84.62\% \\
\quad w/ DeepSeek-V3.2 & 23 & 74.19\% & 15 & \textbf{100.00\%} & 5 & \textbf{83.33\%} & 45 & 86.54\% & 88 & 84.62\% \\
\bottomrule
\end{tabular}
\begin{tablenotes}
\footnotesize
\item \textit{Note:} REX and AdvSCanner entries are taken from their original papers because their implementations are unavailable. AdvSCanner only targets reentrancy vulnerabilities. ``--'' indicates data not available or not applicable.
\end{tablenotes}
\end{threeparttable}
\end{table*}

\subsubsection{Baseline Selection}

We selected three representative LLM-based AEG baselines, covering specialized, general-purpose, and agentic coding paradigms:

\begin{itemize}[leftmargin=*, nosep]
    \item REX~\cite{xiao2025prompt}: A general-purpose AEG framework that leverages intrinsic LLM reasoning capabilities with the Foundry testing stack for end-to-end exploit generation across diverse vulnerability types.
    \item AdvSCanner~\cite{wu2024advscanner}: A specialized tool for reentrancy vulnerabilities that uses static analysis to extract attack flows and relies on hardcoded attack templates with Chain-of-Thought~\cite{wei2022chain, kojima2022large} prompts.
    \item Claude Code~\cite{claude-code}: An agentic coding framework baseline developed by Anthropic~\cite{anthropic}.
\end{itemize}

\subsubsection{Evaluation Metric}

To ensure a rigorous and consistent evaluation, we define a strict \textit{success criterion} that an exploit must satisfy all of the following conditions:

\begin{enumerate}[leftmargin=*]
    \item Test Execution: The generated test case passes \texttt{forge test -vvvv} without runtime errors or assertion failures, and actually executes at least one vulnerable path, rather than merely printing log information.
    \item Constraint Compliance: The generated exploit satisfies the three syntactic constraints defined in Section~\ref{subsec:formal-generator}: version compatibility ($\mathcal{C}_{ver}$), minimalism ($\mathcal{C}_{min}$), and oracle embedding ($\mathcal{C}_{oracle}$).
    \item Target Integrity: The vulnerable contract $C_{vul}$ remains unmodified throughout the testing process.
\end{enumerate}

Since REX and AdvSCanner are not open-sourced, we therefore report the success rates from their original papers on comparable SmartBugs-Curated subdatasets: REX on the same 104-contract benchmark across four vulnerability categories, and AdvSCanner on the same reentrancy subset. We treat these numbers as contextual evidence rather than a fully controlled head-to-head comparison, because their success criteria, LLM backbones, prompts, execution budgets, environments, and manual inspection procedures may differ from ours, and their failure cases cannot be independently rechecked. For Claude Code, we run the baseline on the same 104-contract benchmark and apply the same success criterion as \KASS{}. To investigate the impact of LLMs' built-in knowledge on AEG tasks, we also execute a pure prompting baseline, relying solely on the model's intrinsic capabilities without external knowledge augmentation. We manually check the success of exploit candidates reported. For each case, three authors independently inspected the resulting PoCs and the logs.

\subsubsection{Implementation}

Our agent implementation is built upon Claude Code's sub-agent~\cite{cc-subagent} architecture. The Solodit knowledge base is accessed via the search-solodit-mcp tool~\cite{solodit-mcp}.
We evaluate both \KASS{} and the pure prompting baseline with three representative LLMs: GPT-5.1~\cite{gpt51}, Gemini-3-Flash~\cite{gemini3}, and DeepSeek-V3.2~\cite{ds32}. Moreover, the Claude Code baseline uses DeepSeek-V3.2.
And the experiments were conducted on the Ubuntu 22.04.5 LTS operating system, with an i5-13400 CPU, 64GB of memory.

\subsection{RQ1 (Intrinsic LLM Limitation)}

To assess whether LLMs can perform AEG tasks using their pre-trained knowledge alone, we evaluate a pure prompting baseline that directly generates attack contracts without iterative refinement or external knowledge augmentation. This baseline receives the same vulnerable contract and metadata as the Planner Agent, relying entirely on knowledge acquired during pre-training.

As shown in Table~\ref{tab:comparison}, pure prompting yields limited success: GPT-5.1 performs best at 33.65\%, followed by DeepSeek-V3.2 (9.62\%) and Gemini-3-Flash (4.81\%). The results are also highly uneven across vulnerability types. GPT-5.1 succeeds on 73.33\% of arithmetic cases but only 12.90\% of reentrancy cases, suggesting that simple boundary-condition bugs are better represented in pre-training than complex exploit flows. Safety constraints further limit direct exploit generation: GPT-5.1 refuses 50.96\% of cases, with refusals concentrated in reentrancy (23 of 31) rather than arithmetic (1 of 15). Excluding refusals, its success rate rises to 68.63\%, indicating that capability exists but is difficult to elicit reliably through direct prompting. Gemini-3-Flash additionally suffers from prompt-adherence issues, often generating Foundry-specific components despite the requirement for standalone Solidity attack contracts.

\begin{tcolorbox}[colback=gray!10, colframe=gray!50, boxrule=0.5pt, arc=0mm, left=2pt, right=2pt, top=2pt, bottom=2pt, breakable]
\textbf{Answer to RQ1:} Pre-trained knowledge alone is insufficient for smart contract AEG tasks: the best pure prompting baseline reaches only 33.65\% overall and remains highly sensitive to vulnerability type, safety refusals, and prompt adherence, motivating \KASS{}'s knowledge-augmented and iterative design.
\end{tcolorbox}

\subsection{RQ2 (Effectiveness \& Efficiency)}

RQ2 evaluates \KASS{} from three perspectives: its effectiveness relative to reproduced and previously reported baselines, its runtime and token efficiency, and the remaining failure cases that reveal current limitations.

\textbf{Comparison with Baselines.}
We compare \KASS{} with a same-protocol Claude Code baseline and use the published REX and AdvSCanner results as contextual references because their implementations are unavailable. As shown in Table~\ref{tab:comparison}, \KASS{} with GPT-5.1 achieves a 94.23\% overall success rate, higher than the reproduced Claude Code baseline by 74.04 pp and higher than the reported REX result by 44.23 pp on the comparable 104-contract benchmark. It also reaches 93.55\% on the reentrancy subset, compared with the reported 80\% AdvSCanner result. These REX and AdvSCanner comparisons should be interpreted cautiously, since differences in success criteria, LLM backbones, prompts, budgets, environments, and manual inspection procedures may affect direct comparability.

Qualitatively, \KASS{} differs from these baselines by using retrieval-augmented planning, role-decomposed PoC generation, and exploit-specific validation. Unlike AdvSCanner's reentrancy-specific handcrafted templates and REX's single-pass exploit synthesis, \KASS{} retrieves contract-specific audit knowledge and separates attack planning from executable test generation. This design helps cover all four vulnerability categories while grounding generated PoCs in explicit attack objectives and postconditions.

Compared with Claude Code, \KASS{} adds exploit-specific intent and validation to a general coding workflow. Claude Code can edit files and react to build or test feedback, but without a vulnerability-grounded planning stage it often produces shallow harnesses that deploy contracts or print logs without executing the vulnerable path or asserting attack success. \KASS{} mitigates this failure mode by requiring a structured exploit flow from the Planner Agent and exploit-specific postconditions from the Tester Agent. Even with the same DeepSeek-V3.2 backbone, \KASS{} improves over Claude Code from 20.19\% to 84.62\%.

\textbf{Efficiency Analysis.}
\begin{figure}[t]
  \centering
  \includegraphics[width=0.9\linewidth]{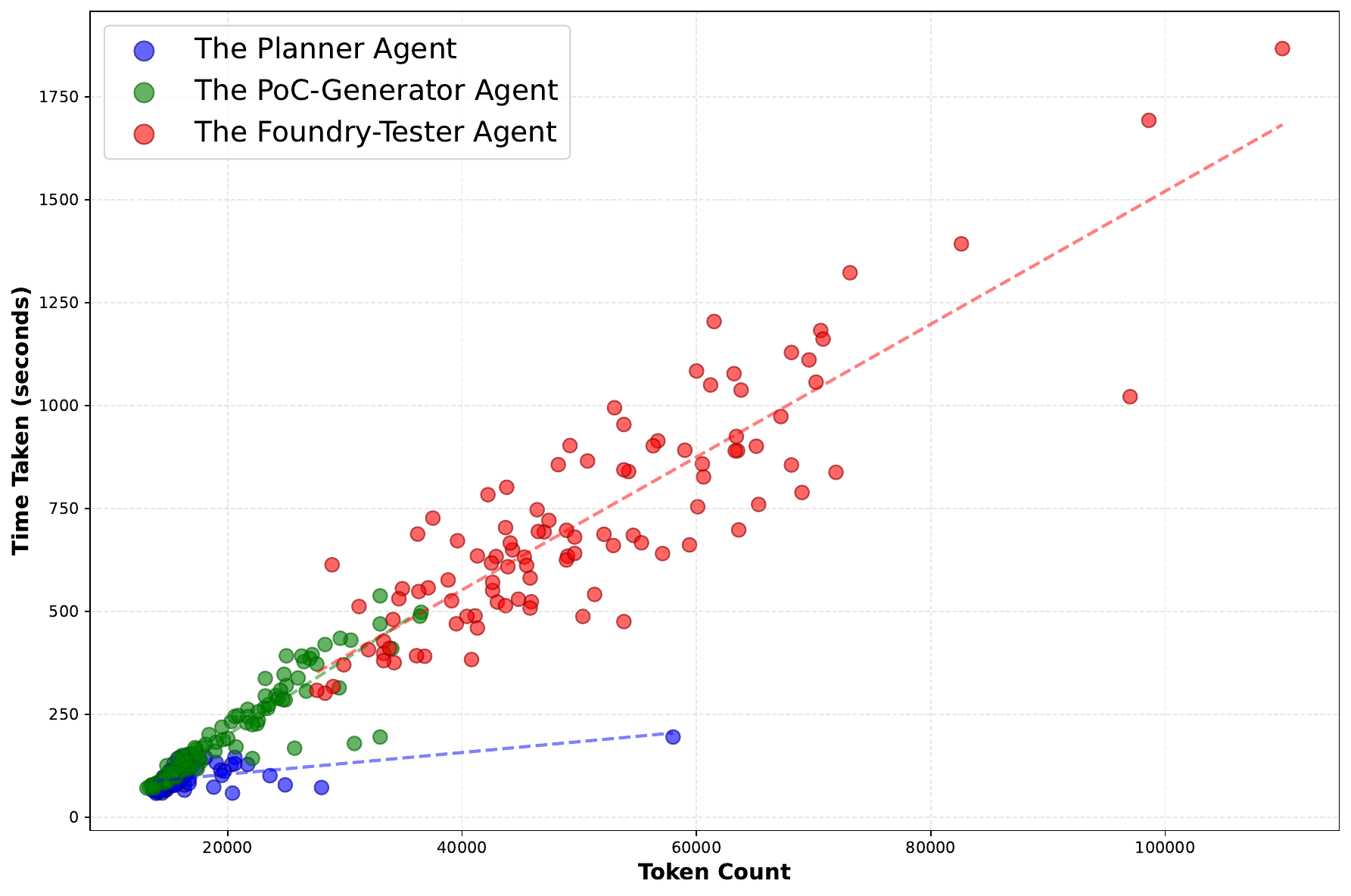}
  \caption{Cost Tokens vs Time.}
  \label{fig:token-vs-time}
\end{figure}
Figure~\ref{fig:token-vs-time} presents the token consumption and time cost distributions across the three agents. The Planner Agent is the most lightweight, consuming an average of 16,370 tokens in 95.1 seconds, reflecting its focused task of generating a structured attack plan from retrieved knowledge.
The PoC-Generator Agent requires moderately more resources (20,241 tokens, 204.5s on average) as it translates abstract plans into executable Solidity code. The Foundry-Tester Agent dominates the overall cost, averaging 50,669 tokens and 724.7 seconds, with a maximum of 110k tokens and 1,867.4 seconds for complex cases. This is expected, as the Tester Agent iteratively executes, analyzes traces, and refines code across multiple inner-loop and outer-loop cycles.

The Pearson correlation~\cite{hauke2011comparison} between token count and execution time is moderate for the Planner Agent ($r$=0.486, $p$<0.001) but strong for the PoC-Generator ($r$=0.917, $p$<0.001) and Foundry-Tester ($r$=0.877, $p$<0.001) agents. The weaker correlation for the Planner Agent reflects its reliance on external retrieval latency, which introduces time overhead independent of generation length.

\textbf{Failure Case and Limitation.} To characterize \KASS{}'s capability boundary, we examine a representative failure involving the \texttt{WALLET} contract~\cite{wallet-contract}. As shown in Listing~\ref{lst:wallet}, the contract contains a reentrancy pattern in \texttt{Collect}, but exploitability is guarded by a temporal constraint: deposits invoke \texttt{Put(0)}, which sets \texttt{acc.unlockTime} to at least the current block timestamp, while \texttt{Collect} requires \texttt{now > acc.unlockTime}. The Planner Agent identified the reentrancy pattern but generated an atomic attack flow that deposited and withdrew in the same transaction context, causing the temporal guard to fail. A successful exploit would require explicit state manipulation, such as inserting \texttt{vm.warp(block.timestamp + 1)} between preparation and interaction. This case shows that \KASS{} remains less reliable when exploitability depends on subtle state-dependent temporal semantics.

\begin{lstlisting}[language=Solidity, caption={Key Functions of the WALLET Contract.}, float=t, label={lst:wallet}]
function Put(uint _unlockTime) public payable{
    var acc = Acc[msg.sender];
    acc.balance += msg.value;
    acc.unlockTime = _unlockTime > now ? _unlockTime : now;
}
function Collect(uint _am) public payable{
    var acc = Acc[msg.sender];
    if( acc.balance >= MinSum && acc.balance >= _am && now > acc.unlockTime){
        // REENTRANCY
        if(msg.sender.call.value(_am)()){
            acc.balance -= _am;
        }
    }
}
function() public payable{
    Put(0);
}
\end{lstlisting}

\begin{tcolorbox}[colback=gray!10, colframe=gray!50, boxrule=0.5pt, arc=0mm, left=2pt, right=2pt, top=2pt, bottom=2pt, breakable]
\textbf{Answer to RQ2:} \KASS{} with GPT-5.1 achieves a 94.23\% success rate. This is higher than our same-protocol Claude Code baseline and higher than previously reported REX and AdvSCanner results on comparable SmartBugs-Curated subsets, although direct comparability with REX and AdvSCanner is limited by unavailable implementations and potentially different evaluation protocols. The gains come from retrieval-augmented planning, role-decomposed generation, and exploit-specific postcondition checking. Its main cost lies in iterative testing, and its main remaining limitation is reasoning about subtle state-dependent temporal constraints.
\end{tcolorbox}

\subsection{RQ3 (Ablation \& Sensitivity Analysis)}

RQ3 evaluates the contribution of each refinement loop through ablation and examines \KASS{}'s robustness to the choice of LLM backbone.

\begin{table}[tbp]
    \centering
    \caption{Average inner-loop iterations per vulnerability type and ablation of the inner-loop mechanism. SR\textsubscript{w/o IL} denotes success rate without inner-loop refinement.}
    \label{tab:inner-iterations}
    \small
    \setlength{\tabcolsep}{4pt}
    \resizebox{\columnwidth}{!}{%
    \begin{tabular}{@{}lc|ccccc@{}}
        \toprule
        \textbf{Config.} & \textbf{SR\textsubscript{w/o IL}} & \textbf{Reent.} & \textbf{Arith.} & \textbf{DoS} & \textbf{Uncheck.} & \textbf{Avg.} \\
        \midrule
        \KASS{} w/ GPT        & 22.4\% & 2.11 & 2.17 & 1.50 & 2.64 & 2.34 \\
        \KASS{} w/ Gemini & 68.7\% & 1.50 & 1.31 & 1.40 & 1.18 & 1.32 \\
        \KASS{} w/ DeepSeek  & 30\% & 2.07 & 2.29 & 1.88 & 3.00 & 2.06 \\
        \bottomrule
    \end{tabular}
    }
\end{table}

\begin{table}[tbp]
    \centering
    \caption{Impact of outer-loop iterations on success rate.}
    \label{tab:outer-loop}
    \small
    \begin{tabular}{@{}lc|cc@{}}
        \toprule
        \textbf{Config.} & \textbf{w/o Outer-Loop} & \textbf{$K$=1} & \textbf{$K$=2} \\
        \midrule
        \KASS{} w/ GPT        & 78.85\% & 89.42\% & \textbf{94.23\%} \\
        \KASS{} w/ Gemini     & 48.08\% & 66.35\% & \textbf{84.62\%} \\
        \KASS{} w/ DeepSeek   & 65.38\% & 77.88\% & \textbf{84.62\%} \\
        \bottomrule
    \end{tabular}
\end{table}

\textbf{Inner-Loop Ablation.}
Table~\ref{tab:inner-iterations} reports the average inner-loop iterations and ablation result (SR\textsubscript{w/o IL}). Removing the inner loop sharply reduces GPT-5.1 from 94.23\% to 22.4\% ($-$71.83 pp) and DeepSeek-V3.2 from 84.62\% to 30\% ($-$54.62 pp), showing that code-level refinement is critical. Most exploits converge within 2--3 cycles (avg. 1.32--2.34 iterations).
Gemini-3-Flash is an exception: its SR\textsubscript{w/o IL} remains 68.7\% because it often generates cleaner first-pass code, but its weaker initial strategies require outer-loop compensation.

\textbf{Outer-Loop Ablation.}
Table~\ref{tab:outer-loop} shows consistent gains as $K$ increases from 0 to 2, confirming the value of strategic replanning beyond code correction. Gemini-3-Flash benefits most (+36.54 pp, 48.08\% $\to$ 84.62\%), GPT-5.1 least (+15.38 pp, 78.85\% $\to$ 94.23\%), and DeepSeek-V3.2 increases steadily (+19.24 pp, 65.38\% $\to$ 84.62\%).
Some Gemini failures stem from instruction non-compliance, especially upgrading the contract's Solidity version and violating the Version Compatibility constraint ($\mathcal{C}_{ver}$) in Section~\ref{subsec:generater}.

\textbf{LLM Backbone Sensitivity.}
\KASS{} achieves 84.62\%--94.23\% across all three backbones, indicating robustness to backbone choice. GPT-5.1 achieves the highest success rate but relies heavily on inner-loop correction; Gemini-3-Flash produces cleaner first-pass code but depends on strategic replanning; DeepSeek-V3.2 provides a balanced cost-performance alternative.

\begin{tcolorbox}[colback=gray!10, colframe=gray!50, boxrule=0.5pt, arc=0mm, left=2pt, right=2pt, top=2pt, bottom=2pt, breakable]
\textbf{Answer to RQ3:} Both loops are necessary: removing the inner loop costs up to 71.83 pp, and removing the outer loop costs up to 36.54 pp. \KASS{} remains robust across backbones, with GPT-5.1 favoring code correction, Gemini strategic replanning, and DeepSeek-V3.2 balancing both.
\end{tcolorbox}

\subsection{RQ4 (Real-World CVE Validation)}

RQ4 evaluates whether \KASS{} can reproduce vulnerabilities from real-world CVE disclosures, where descriptions are often less structured than benchmark annotations. We run \KASS{} with DeepSeek-V3.2 on the 11-contract CVE dataset and judge each output using the same four-part success criterion.

\begin{table}[tbp]
    \centering
    \caption{Results on real-world CVE-tagged vulnerable contracts.}
    \label{tab:cve-results}
    \scriptsize
    \setlength{\tabcolsep}{2pt}
    \begin{tabularx}{\linewidth}{@{}lrlcX@{}}
        \toprule
        \textbf{CVE ID} & \shortstack{\textbf{Funcs.}} & \textbf{Type} & \textbf{Pass} & \textbf{PoC Outcome} \\
        \midrule
        CVE-2019-15080 & 15 & ERC20 Token & \checkmark & Owner takeover; mint and blacklist \\
        CVE-2020-17752 & 25 & ERC20 Token & $\times$ & - \\
        CVE-2020-17753 & 75 & Crowdsale & \checkmark & Whitelist bypass; token theft \\
        CVE-2020-35962 & 43 & Protocol Fee Vault & \checkmark & Unauthorized vault drain \\
        CVE-2021-3004 & 85 & ERC20 Token & \checkmark & Recipient balance zeroed \\
        CVE-2021-33403 & 33 & ERC20 Token & $\times$ & - \\
        CVE-2021-34272 & 18 & ERC20 Token & \checkmark & Owner takeover; mint and freeze \\
        CVE-2021-34273 & 16 & ERC20 Token & \checkmark & Owner takeover \\
        CVE-2024-51424 & 14 & ERC20 Token & \checkmark & Owner takeover; token allocation \\
        CVE-2024-51425 & 27 & ERC20 Token & \checkmark & Owner takeover; supply capture \\
        CVE-2025-56207 & 77 & ERC721 NFT & \checkmark & Permanent NFT burn \\
        \bottomrule
    \end{tabularx}
\end{table}

\textbf{Results Overview.}
As shown in Table~\ref{tab:cve-results}, \KASS{} successfully validates 9 of 11 CVE-tagged contracts. This result is lower than \KASS{}'s best SmartBugs-Curated performance but remains strong given the additional noise in real-world disclosures, including incomplete vulnerability descriptions, heterogeneous contract styles, and cases where the exploit objective must be inferred from sparse CVE text.

The PoC Outcome column summarizes the concrete security consequence demonstrated or targeted by each generated proof-of-concept. This shows that \KASS{} goes beyond conventional vulnerability detection and textual descriptions by making the practical impact and consequences of each vulnerability more explicit and intuitive.

\textbf{Failure Analysis.}
The two failures stem from tests that passed without exercising the vulnerable path. For CVE-2021-33403, the test only initialized balances and asserted an overflow condition, without calling the vulnerable function. For CVE-2020-17752, the test only reasoned about MON's arithmetic behavior, without invoking the payable purchase or minting logic or checking the resulting contract state.

\begin{tcolorbox}[colback=gray!10, colframe=gray!50, boxrule=0.5pt, arc=0mm, left=2pt, right=2pt, top=2pt, bottom=2pt, breakable]
\textbf{Answer to RQ4:} The CVE experiment shows that \KASS{} performs well on diverse and complex real-world vulnerable contracts, demonstrating effectiveness beyond curated benchmark datasets while using executable tests to confirm vulnerable paths and attack outcomes.
\end{tcolorbox}

\subsection{RQ5 (Exploitability Assessment)}

Detection-centric tools still struggle to assess the practical severity of reported vulnerabilities and separate exploitable flaws from false positives. RQ5 examines whether \KASS{}'s structured outputs help address these limitations through two representative cases.

\textbf{Case 1: Exploit Generation with Damage Quantification.}
Beyond success rate metrics, \KASS{}'s structured attack plans document exploitation flows and quantify potential damage.
For instance, when analyzing the BEC Token~\cite{bec-token-contract} integer overflow vulnerability, \KASS{} generated a plan~\cite{bec-exploit-plan} with a 4-step preparation phase and a 6-step interaction process.
Its \textit{Post-Attack State} analysis further shows that the attacker loses only gas fees, each receiver gains $2^{255}$ BEC tokens (approximately $5.79 \times 10^{76}$), the supply inflates by $2^{256}$ tokens, and the token economy is destroyed by hyperinflation.
Thus, \KASS{} turns an \textit{overflow detected} warning into an explicit damage assessment.
The generated exploit was verified by Foundry under the \textit{Overflow to 0} attack vector, setting \texttt{\_value} $= 2^{255}$ and \texttt{cnt} $= 2$. Since $2 \times 2^{255} = 2^{256}$ wraps to 0, the balance check passes and no balance is deducted from the attacker, while each receiver is credited $2^{255}$ tokens from thin air.

\textbf{Case 2: Semantic False Positive Filtering.}
\KASS{} also shows potential as a post-analysis false positive filter. We illustrate this with a case~\cite{slither-fp-contract} where Slither~\cite{feist2019slither} reports a reentrancy vulnerability that is, in practice, unexploitable.
As shown in Listing~\ref{lst:false-positive}, Slither flags this function because state updates occur after the external \texttt{transfer()}, violating the Checks-Effects-Interactions pattern. However, \texttt{transfer()} and \texttt{send()} impose a 2,300 gas stipend, unlike \texttt{call.value()} which forwards all available gas. This makes profitable reentrancy infeasible, but syntactic detectors may still report the CEI violation without modeling the EVM gas constraint~\cite{zheng2023turn}.
When we fed this contract to \KASS{} using Slither's vulnerability label as input metadata, the Planner Agent identified the 2,300-gas limitation as a necessary precondition during planning.
This precondition captures the practical infeasibility of exploitation and suggests that \KASS{} can complement detection pipelines by checking whether reported vulnerabilities are genuinely exploitable.

\begin{lstlisting}[language=Solidity, caption={A false positive reported by Slither as a reentrancy vulnerability.}, float=t, label={lst:false-positive}]
function buy_fromContract() payable
  public returns (uint256 _amount_) {
    require (msg.value >= 0);
    _amount_ = msg.value / buyPrice;
    if (_amount_ > balances[this]) {
        _amount_ = balances[this];
        uint256 valueWei = _amount_ * buyPrice;
        msg.sender.transfer(msg.value - valueWei);
    }
    balances[msg.sender] += _amount_;
    balances[this] -= _amount_;
    Transfer(this, msg.sender, _amount_);
    return _amount_;
}
\end{lstlisting}

\begin{tcolorbox}[colback=gray!10, colframe=gray!50, boxrule=0.5pt, arc=0mm, left=2pt, right=2pt, top=2pt, bottom=2pt, breakable]
\textbf{Answer to RQ5:} \KASS{}'s structured outputs go beyond binary detection by linking reported vulnerabilities to executable attack evidence. They document exploitation flows, quantify concrete damage, and help identify unexploitable patterns such as gas-limited \texttt{transfer()} reentrancy, making them useful for both impact assessment and semantic false positive filtering.
\end{tcolorbox}

\section{Discussion}
\label{sec:discuss}

\subsection{Related Works}
\label{sec:related-work}

Smart contract vulnerability detection has evolved from static analyzers~\cite{feist2019slither, tikhomirov2018smartcheck, tsankov2018securify, brent2020ethainter} and symbolic execution engines~\cite{luu2016making, mossberg2019manticore, mythril2026, frank2020ethbmc, bose2022sailfish} to coverage-guided fuzzers~\cite{jiang2018contractfuzzer, grieco2020echidna, choi2021smartian, shou2023ityfuzz, wustholz2020harvey} and LLM-driven auditing frameworks~\cite{sun2024gptscan, liu2025propertygpt, wei2025smartauditflow, wei2025advanced, ma2025combining}.
Besides, LLMs have also been widely studied for code generation through structured prompting, self-planning, self-collaboration, multi-agent decomposition, and tool-integrated repository-level coding~\cite{li2025structured, jiang2024self, dong2024self, hong2024metagpt, islam2024mapcoder, zhang2024codeagent}.

Traditional AEG systems~\cite{avgerinos2014automatic, schwartz2011q, xu2022bofaeg, liu2022automated, pewny2019steroids} generate exploits for C/C++ memory vulnerabilities, none of which transfer to smart contract environments. In the smart contract domain, teEther~\cite{krupp2018teether} synthesizes ETH-draining transactions via path-condition analysis but is limited to simple ETH-transfer bugs; AdvSCanner~\cite{wu2024advscanner} targets reentrancy via LLM and static analysis but relies on hardcoded templates; REX~\cite{xiao2025prompt} applies a general-purpose LLM pipeline without grounding in real-world attack semantics; and general coding-agent frameworks such as Claude Code~\cite{claude-code} provide useful development automation but are prone to task drift, often producing simple logging or deployment harnesses rather than actually exercising the vulnerable path.

In contrast, \KASS{} goes beyond vulnerability detection by targeting executable exploit verification. Compared with general LLM-based code generation, smart contract AEG is substantially more complex because it must satisfy specific blockchain requirements. \KASS{} addresses these challenges by integrating real-world audit knowledge, binding attack plans to formal generation and validation constraints, and using a hierarchical dual-loop mechanism to jointly refine code-level errors and strategy-level attack assumptions.

\subsection{Threats to Validity}
\label{sec:threats}

The main internal threat is LLM non-determinism: identical inputs may yield different outputs. We mitigate this by using consistent hyperparameters across experiments.
Baseline comparability is limited because REX and AdvSCanner are not open-sourced; we therefore rely on their published results rather than rerunning them in our environment. As a result, differences in success criteria, LLM backbones, prompts, running budgets, execution environments, manual inspection procedures, and unreproducible failure cases may affect the comparison. We address this limitation by clearly marking REX and AdvSCanner as reported results, restricting the comparison to comparable SmartBugs-Curated subsets, and additionally evaluating Claude Code with DeepSeek-V3.2 on the same benchmark and under the same success criterion as \KASS{}.
External validity is limited by benchmark scope. SmartBugs-Curated covers representative machine-auditable vulnerabilities, and our CVE experiment adds real-world cases, but broader machine-unauditable bugs remain future work.

\section{Conclusion}
\label{sec:conclusion}

We presented \KASS{}, a knowledge-augmented framework that bridges smart contract vulnerability detection and executable exploit verification. \KASS{} relies on three core mechanisms: retrieval-augmented planning over real-world audit knowledge, formal generation and validation constraints that force PoCs to exercise vulnerable paths, and a hierarchical dual-loop that repairs code errors while replanning invalid attack strategies.
These mechanisms allow \KASS{} to achieve a 94.23\% success rate across 104 SmartBugs-Curated contracts, exceeding our same-protocol Claude Code baseline and previously reported REX and AdvSCanner results on comparable benchmark subsets.
On 11 real-world CVE-tagged contracts, \KASS{} successfully validates 9 cases, further showing its ability to transfer beyond curated benchmarks. Beyond exploit generation, \KASS{}'s structured outputs also support damage quantification and semantic false positive filtering for static analysis pipelines.


\bibliographystyle{IEEEtran}
\bibliography{references}

\end{document}